# Fusion of Complex Networks-based Global and Local Features for Texture Classification

Zhengrui Huang
Academy of Digital China (Fujian)
Fuzhou University
Fuzhou, China
zhengruihuang1994@gmail.com

*Abstract*—To realize accurate texture classification, this article proposes a complex networks (CN)-based multi-feature fusion method to recognize texture images. Specifically, we propose two feature extractors to detect the global and local features of texture images respectively. To capture the global features, we first map a texture image as an undirected graph based on pixel location and intensity, and three feature measurements are designed to further decipher the image features, which retains the image information as much as possible. Then, given the original band images (BI) and the generated feature images, we encode them based on the local binary patterns (LBP). Therefore, the global feature vector is obtained by concatenating four spatial histograms. To decipher the local features, we jointly transfer and fine-tune the pre-trained VGGNet-16 model. Next, we fuse and connect the middle outputs of max-pooling layers (MP), and generate the local feature vector by a global average pooling layer (GAP). Finally, the global and local feature vectors are concatenated to form the final feature representation of texture images. Experiment results show that the proposed method outperforms the state-of-the-art statistical descriptors and the deep convolutional neural networks (CNN) models.

## I. Introduction

Recent advancements in computer vision (CV) give help to some applications, including facial expression detection [1], scene recognition [2], texture classification [3], etc. Indeed, the most widely studied part of CV is texture classification, based on which most of paradigms are developed, such as texture synthesis [4]. To realize accurate texture classification, a large number of methods were proposed [5], including the statistical methods, the model-based methods, the learning-based methods, etc. Interestingly, these methods have different abilities of detecting features. As mentioned in [6], the common features are mainly divided into two parts: 1) global features and 2) local features, and different extractors show advantages in different scenarios, i.e., the combination or fusion of multiple extractors outperforms single extractor.

Specifically, the statistical methods are good at extracting global texture features. The widely used statistical descriptor was local binary patterns (LBP) that was first proposed for rotation invariant texture image classification [7]. Following LBP, some LBP-based variants were developed. In [8], a local directional number (LDN)-based descriptor was proposed to distinguish similar structural modes. To reduce the impact of imaging, Ding *et al.* introduced a dual-cross pattern (DCP) for facial expression detection [9]. A local bit-plane decoded pattern (LBDP) was designed for image matching and retrieval [10], and Chakraborty *et al.* proposed centre symmetric quadruple patterns (CSQP) to describe larger neighbourhood around a center pixel [11].

However, it is worth noting that the statistic-based methods can not effectively decipher local information, such as shapes and structures. Fortunately, the model-based methods provide promising solutions. In [12], a CN-based local spatial pattern mapping (CN-LSPM) was introduced to detect shape patterns, and Bhattacharjee *et al.* used a gravitational model to extract object edges [3]. Moreover, the learning-based models gain attention. A CNN architecture named VGGNet was developed to extract local information, and included convolutional layers (Conv), max-pooling layers (MP), and fully connected layers (FC) [13]. Following [13], He *et al.* developed a residual network (ResNet) to ease the training of VGGNet [14], and a densely connected CNN (DenseNet) was proposed to alleviate the problem of gradient vanishing [15]. Recently, Xu *et al.* adopted multi-layer outputs derived from VGGNet to describe local image information [16], Jiang *et al.* combined CNN with Gabor orientation filters (GoF) for texture recognition [17], and a pair-wise difference pooling (PDP)-based bi-linear CNN was proposed to describe detailed texture differences [18].

Motivated by the above, we comprehensively study in this article a new feature descriptor by combining global and local features of texture images, as shown in Fig. 1. Specifically, we propose two feature extractors to capture the global and local information of texture images, respectively:

- Global feature extractor (GFE): GFE consists of two steps: 1) CN mapping (CNM) and 2) LBP encoding. In the first step, we apply CNM on each band of a texture image with the help of pixel distance and intensity, and three CN-based measurements (clustering coefficient (CC), degree centrality (DC), and eigenvector centrality (EC)) are selected to evaluate image features, thus obtaining band images (BI) and feature images over different bands, respectively, as shown in Fig. 2. In the second step, we jointly encode BIs and three feature images based on the uniform LBP (ULBP), and the global feature vector is generated by concatenating four spatial histograms.

- Local feature extractor (LFE): In LFE, we first jointly transfer and fine-tune the pre-trained VGGNet-16 model based on BIs and three feature images, and obtain the outputs of middle layers (MPs) derived from the fine-tuned VGGNet-16 model, where we only transfer the backbone of VGGNet-

This work was supported in part by The National Key Research and Development Project under Grant 2017YFB0504202.

16 model. Second, to capture the natural representation of feature maps, we use a global average pooling layer (GAP) to process the outputs of MPs derived from middle layers. Finally, the local feature vector is obtained by connecting the multi-layer feature vectors.

The remainder of this article is summarized as follows. In Section II, we introduce the proposed method. The results and analyses are shown in Section III, and the conclusions are drawn in Section IV.

*Notations*: Scalars are denoted by italic letters, and vectors and matrices are denoted by bold-face lower-case and bold-face upper-case letters, respectively. For a real $x$, $\mathbb{I}(x)$ denotes the indicator function that equals 1 if $x > 0$, and 0 otherwise. For a set $\mathcal{X}$, $|\mathcal{X}|$ denotes its cardinality.

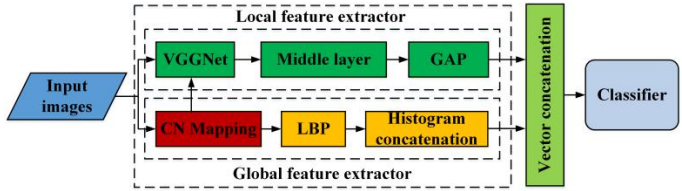

Fig. 1. Overview of the proposed method.

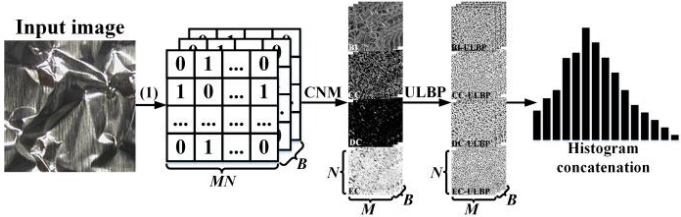

Fig. 2. Overview of GFE.

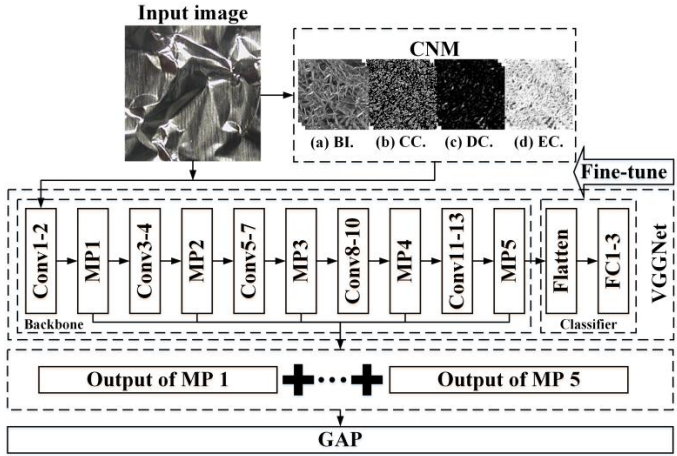

Fig. 3. Overview of LFE.

## II. PROPOSED METHOD

### A. GFE

GFE mainly consists of two steps: 1) CN mapping and 2) LBP encoding. In the first step, we model a multi-band texture image, denoted by $\mathbf{I}_{M \times N \times B}$, as an undirected graph, denoted by $\mathbf{G} = (\mathbf{V}, \mathbf{L})$, where $M \times N$ is the number of pixels in each band, $B$ is the number of bands, $\mathbf{V}$ is the matrix of pixels, and $\mathbf{L}$ is the adjacent matrix, whose entry $l_{i,j}$ equals 1 if pixel $i$ connects with pixel $j$, and 0 otherwise, so the CNM over a band can be expressed as follows:

$$l_{i,j} = \begin{cases} 1, & \text{if } w_{i,j} \leq t \\ 0, & \text{otherwise} \end{cases} \quad (1)$$

where $t$ denotes the similarity threshold, and $w_{i,j}$ denotes the weight of $l_{i,j}$ [12]:

$$w_{i,j} = \begin{cases} \dfrac{d_{i,j}^2 + r^2 |I_i - I_j|/255}{2r^2}, & \text{if } d_{i,j} \leq r \\ 0, & \text{otherwise} \end{cases} \quad (2)$$

where $r$ is the threshold of search-radius, $I$ is the pixel intensity, and $d_{i,j}$ denotes the pixel distance:

$$d_{i,j} = \sqrt{(x_i - x_j)^2 + (y_i - y_j)^2} \quad (3)$$

where $(x, y)$ is the pixel coordinate. Thereby, we can obtain the adjacent matrix $\mathbf{L}$ by traversing all pixels, and $\mathbf{L}$ can be further used to calculate the features of each CN model. Moreover, it is notable that $\mathbf{L}$ is a symmetric matrix, so we only need to traverse the pixels in the upper triangular matrix of $\mathbf{L}$, which effectively speeds up the image processing.

Based on the adjacent matrix $\mathbf{L}$ derived from (1)-(3), we adopt three measurements (CC, DC, and EC) to represent the features of the CN model [19], which effectively retains the image information as much as possible, and use these features to discriminate differences between two CN models:

$$\mathbb{C}(i) = \dfrac{2c_i}{k_i(k_i - 1)} \quad (4)$$

where $c_i$ is the number of edges between neighbor pixels around pixel $i$, i.e., pixels with higher (2) are more likely to be clustered into the same community, namely, the same class, and $k_i$ is the number of edges connected with pixel $i$ [20]:

$$k_i = \sum l_{i,j} = \sum l_{j,i} \quad (5)$$

Then, we define DC based on the form of energy:

$$\mathbb{D}(i) = \left(\dfrac{k_i}{MN - 1}\right)^2 \quad (6)$$

And EC is expressed as the form of entropy:

$$\mathbb{E}(i) = -\left(\lambda \sum_{j=0}^{M^2-1} l_{i,j} u_j\right) \left(\log_2 \left(\lambda \sum_{j=0}^{M^2-1} l_{i,j} u_j\right)\right) \quad (7)$$

where $\lambda$ is the reciprocal of the maximum eigenvalue of $\mathbf{L}$, and $u_j$ is the element of eigenvector derived from $\lambda^{-1}$.

In the second step, following (4)-(7), three feature images are generated, as shown in Fig. 2. To reduce the impact of rotation and imaging, we apply ULBP on BIs and three feature images, respectively:

$$\text{ULBP}(I_c) = \begin{cases} \sum_{p=0}^{P-1} \mathbb{I}(I_p - I_c) 2^p, & \text{if } \mathbb{U}(\text{LBP}(I_c)) \leq 2 \\ P+1, & \text{otherwise} \end{cases} \quad (8)$$

where $I_c$ and $I_p$ denote the intensity of a central pixel $c$ and a neighbour pixel $p$, respectively, $P$ is the number of neighbour pixels, $\text{LBP}(I_c) = \sum_{p=0}^{P-1} \mathbb{I}(I_p - I_c) 2^p$, and $\mathbb{U}(\cdot)$ denotes the uniform pattern function [7]:

$$\mathbb{U}(\text{LBP}(I_c)) = |\mathbb{I}(I_{P-1} - I_c) - \mathbb{I}(I_0 - I_c)| + \sum_{p=1}^{P-1} |\mathbb{I}(I_p - I_c) - \mathbb{I}(I_{p-1} - I_c)| \quad (9)$$

Finally, the global feature vector is obtained by connecting four spatial histograms:

$$\mathbf{f}_{\text{global}} = [\mathcal{H}_{\text{BI-ULBP}}, \mathcal{H}_{\text{CC-ULBP}}, \mathcal{H}_{\text{DC-ULBP}}, \mathcal{H}_{\text{EC-ULBP}}] \quad (10)$$

where $\mathcal{H}$ is the spatial histogram, and $|\mathbf{f}_{\text{global}}| = B \times 59 \times 4$.

### B. LFE

In LFE, given the generated set of BIs and three feature images, denoted by $[\mathbf{S}_{\text{BI}}, \mathbf{S}_{\text{CC}}, \mathbf{S}_{\text{DC}}, \mathbf{S}_{\text{EC}}]$, we jointly transfer and fine-tune the pre-trained VGGNet-16 model to extract local image information, as shown in Fig. 3.

Specifically, we first transfer the backbone the pre-trained VGGNet-16 model that consists of 13 Convs and 5 MPs, where the kernel sizes of Conv and MP are $3 \times 3$ and $2 \times 2$, respectively, the activation function is the rectified linear unit (ReLU), and the stride size is 2×2. Second, we fine-tuned the VGGNet-16 model by updating the model weights of the backbone. Next, to better describe the local features, we tend to fuse the outputs of middle layers (MP) of the fine-tuned VGGNet-16 model, where the depths of five MPs are equal to 64, 128, 256, 512, and 512, respectively. To jointly make the feature maps corresponding to the number of bands and reduce the length of the generated final feature vector, the GAP is selected to process the middle outputs, and the output of each MP can be expressed as follows:

$$\mathbf{M}_l = \mathbb{M}([\mathbf{S}_{\text{BI}}, \mathbf{S}_{\text{CC}}, \mathbf{S}_{\text{DC}}, \mathbf{S}_{\text{EC}}]), l \in \mathcal{L} = \{0, ..., L\} \quad (11)$$

where $\mathbb{M}(\cdot)$ denotes a MP, and $L$ is the number of MPs.

Finally, we connect the outputs of multiple MPs based on (11) and obtain the local feature vector by global averaging:

$$\mathbf{f}_{\text{local}} = [\mathbb{G}(\mathbf{M}_0), ..., \mathbb{G}(\mathbf{M}_l)], l \in \mathcal{L} = \{0, ..., L\} \quad (12)$$

where $\mathbb{G}(\cdot)$ denotes a GAP, and $|\mathbf{f}_{\text{local}}| = 5 \times (2^6 + ... + 2^{6+l})$. As a result, we concatenate (10) and (12) to generate the final feature vector, denoted by $[\mathbf{f}_{\text{global}}, \mathbf{f}_{\text{local}}]$.

## III. EXPERIMENTS AND ANALYSES

### A. Datasets Description and Environmental Settings

In this article, we select three datasets to verify the effectiveness of our proposed method, including CUReT [21], KTH-TIPS2-b [22], and Outex-TC-00013 [23]:

- CUReT: The dataset contains 61 categories, and a category has 205 samples with a size of 640×480 pixels.
- KTH-TIPS2-b: The 11 materials are placed in separate files, and each files contains 4 dictionaries, where each dictionary includes 108 images (200×200 pixels).
- Outex-TC-00013: This dataset contains 1360 samples that are divided into 68 classes, and the size is 128×128 pixels.

The parameters of GFE are given by: $r=3$ and $t=0.315$, and LFE is jointly transferred and fine-tuned, where the number of epochs and the batchsize are 10 and 32, respectively, the optimizer is root mean square prop (RMSProp), and the learning rate and the discounting factor equal 0.001 and 0.9, respectively. As shown in Fig .1, the support vector machine (SVM) is selected for image classification, where the kernel function is linear, and the penalty coefficient and the kernel coefficient equal 1.0 and $1/|[\mathbf{f}_{\text{global}}, \mathbf{f}_{\text{local}}]|$, respectively, and the overall accuracy (OA) is used to evaluate the differences between the predicted and true labels. Moreover, before fine-tuning our model, the images in datasets are reshaped as 128×128 pixels and randomly divided into three sets: 1) fine-tune the transferred VGGNet-16 model (20%) 2) train the SVM (50%) and 3) test the SVM (30%). It is notable that CUReT, KTH-TIPS2-b, and Outex-TC-00013 all consist of natural texture images, and all results are averaged over a number of independent experiments through Python, including KERAS, OPENCV, NETWORKX, etc.

### B. Comparisons and Analyses

In this section, the statistical descriptors and the deep CNN (DCNN) models are compared with the proposed method:

- Statistical descriptors: LBP, local derivative patterns (LDP) [24], local energy patterns (LEP) [25], LDN, DCP, dominant rotated LBP (DRLBP) [26], CSQP, local directional ternary patterns (LDTP) [27], local concave-and-convex micro-structure patterns (LCCMSP) [28], local neighborhood difference patterns (LNDP) [29], and local concave micro structure patterns (LCvMSP) [30].
- DCNN models: VGGNet-16, ResNet-50, DenseNet-201, depthwise separable convolutions (DSC) [31], NASNet-Large [32], MF$^2$Net, GCN, BCNN-PDP, GCNN, hybrid deep features (HDF) [33], and parallel neural networks (PNN).

As shown in Table Ⅰ, we present the OAs of the proposed method, the statistical descriptors, and the DCNN models on three datasets, respectively, and can find that the OAs of our method outperform those of statistical descriptors and DCNN models, where the highest OAs on three datasets reach

97.48%, 99.39%, and 88.99%, respectively. Moreover, to further verify the effectiveness of our method, we compute and present the confusion matrices on three datasets, and the OAs of most classes reach 100%, as shown in Fig. 4.

TABLE I
OAs (%) of The Proposed Method, The Statistical Descriptors, and The DCNN Models

| | Method | CUReT | KTH-TIPS2-b | Outex-TC-00013 |
|---|---|---|---|---|
| | LBP | 91.03 | 89.63 | 77.97 |
| | LDP | 85.89 | 82.05 | 76.99 |
| | LEP | 77.03 | 76.50 | 73.00 |
| | LDN | 84.67 | 81.32 | 74.04 |
| | DCP | 87.86 | 77.76 | 70.83 |
| ▲ | DRLBP | 91.29 | 91.74 | 77.55 |
| | CSQP | 89.85 | 82.69 | 76.20 |
| | LDTP | 92.25 | 90.97 | 80.85 |
| | LCCMSP | 94.92 | 93.51 | 84.78 |
| | LNDP | 91.64 | 87.85 | 77.16 |
| | LCvMSP | 96.44 | 96.10 | 88.97 |
| | VGGNet-16 | 86.58 | 92.31 | 83.01 |
| | ResNet-50 | 84.72 | 95.62 | 78.82 |
| | DenseNet-201 | 87.08 | 96.17 | 79.12 |
| | DSC | 84.36 | 94.29 | 75.26 |
| | NASNet-Large | 85.45 | 94.78 | 79.82 |
| ▼ | MF²Net | 92.56 | 97.26 | 86.03 |
| | GCN | 93.76 | 98.72 | 88.33 |
| | BCNN-PDP | 96.27 | 98.02 | 86.72 |
| | GCNN | 95.25 | 97.88 | 80.26 |
| | HDF | 87.77 | 91.67 | 86.51 |
| | PNN | 91.38 | 94.56 | 84.70 |
| | **Our method** | **97.48** | **99.65** | **88.99** |

▲: Statistical descriptors ▼: DCNN models

To further improve the performance of the proposed method, we combine our method with the dimension reducing (DC) methods, as shown in Table Ⅱ, including principal components analysis (PCA), linear discriminant analysis (LDA), and chi- square test (Chi2), which jointly measures the importance of features and removes the noises of spatial histograms. In this article, the principle components are determined based on the curve of cumulative explained variance, as shown in Fig. 4, and the feature numbers of LDA and Chi2 are both set to the number of classes. As demonstrated in Table Ⅱ, we can know that PCA gives the best performance, where the length of original feature vector is 6596, and the optimal numbers of principal components in three datasets are equal to 1708, 665, and 761, respectively.

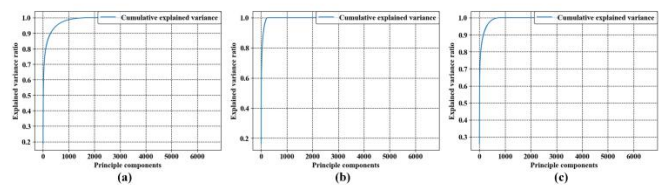

Fig. 5. Cumulative explained variances. (a) CUReT. (b) KTH-TIPS2-b. (c) Outex-TC-00013.

TABLE II
OAs (%) of The Proposed Method with Different DC Methods

| Method | CUReT | KTH-TIPS2-b | Outex-TC-00013 |
|---|---|---|---|
| PCA | 97.95 | 100.00 | 89.18 |
| LDA | 94.26 | 97.90 | 84.68 |
| Chi2 | 91.53 | 97.65 | 81.51 |

IV. CONCLUSION

This article proposed a new texture descriptor by combining the global and local features of texture images. Specifically, in GFE, the global features were extracted based on CNM and LBP encoding, and the global feature vector was obtained by histogram concatenation. In LFE, we jointly transferred and fine-tuned the VGGNet-16 model, and the local feature vector was generated by globally fusing and averaging the outputs derived from middle layers. By fusing the generated global and local features, our proposed method outperformed the state-of- the-art statistic-based descriptors and DCNN models, thus making great contributions to texture classification. However, there are problems not addressed in this work, which are worth being discussed in the future. For instance, it is interesting to further fuse hidden features from multi-scales and multi-layers.

REFERENCES

[1] Z. Xi, Y. Niu, J. Chen, X. Kan, and H. Liu, "Facial Expression Recognition of Industrial Internet of Things by Parallel Neural Networks Combining Texture Features," IEEE Transactions on Industrial Informatics, vol. 17, no. 4, pp. 2784-2793, 2021.
[2] Z. Huang, "Multi-Feature Fusion-based Scene Classification Framework for HSR Images," arXiv e-prints, p. arXiv:2105.10758, 2021.
[3] D. Bhattacharjee and H. Roy, "Pattern of Local Gravitational Force (PLGF): A Novel Local Image Descriptor," IEEE Transactions on Pattern Analysis and Machine Intelligence, vol. 43, no. 2, pp. 595-607, 2021.
[4] Z. Huang, X. Lin, and C. Chen, "Fast Texture Synthesis for Discrete Example-Based Elements," IEEE Access, vol. 8, no. 1, pp. 76683-76691, 2020.
[5] A. Humeau-Heurtier, "Texture feature extraction methods: A survey," IEEE Access, vol. 7, pp. 8975-9000, 2019.
[6] D. Kumar and D. Sharma, "Multi-modal Information Extraction and Fusion with Convolutional Neural Networks," in 2020 International Joint Conference on Neural Networks (IJCNN), 2020, pp. 1-9.
[7] T. Ojala, M. Pietikainen, and T. Maenpaa, "Multiresolution gray-scale and rotation invariant texture classification with local binary patterns," IEEE Transactions on Pattern Analysis and Machine Intelligence, vol. 24, no. 7, pp. 971-987, 2002.
[8] A. R. Rivera, J. R. Castillo, and O. O. Chae, "Local Directional Number Pattern for Face Analysis: Face and Expression Recognition," IEEE Transactions on Image Processing, vol. 22, no. 5, pp. 1740-1752, 2013.
[9] C. Ding, J. Choi, D. Tao, and L. S. Davis, "Multi-Directional Multi-Level Dual-Cross Patterns for Robust Face Recognition," IEEE Transactions on Pattern Analysis and Machine Intelligence, vol. 38, no. 3, pp. 518-531, 2016.
[10] [10] S. R. Dubey, S. K. Singh, and R. K. Singh, "Local Bit-Plane Decoded Pattern: A Novel Feature Descriptor for Biomedical Image Retrieval," IEEE Journal of Biomedical and Health Informatics, vol. 20, no. 4, pp. 1139-1147, 2016.
[11] S. Chakraborty, S. K. Singh, and P. Chakraborty, "Centre symmetric quadruple pattern: A novel descriptor for facial image recognition and retrieval," Pattern Recognition Letters, vol. 115, pp. 50-58, 2018/11/01/ 2018.

[12] S. Thewsuwan and K. Horio, "Texture classification by local spatial pattern mapping based on complex network model," Int. J. Innov. Comput. Inf. Control, vol. 14, no. 3, pp. 1113-1121, 2018.

[13] K. Simonyan and A. Zisserman, "Very deep convolutional networks for large-scale image recognition," arXiv preprint arXiv:1409.1556, 2014.

[14] K. He, X. Zhang, S. Ren, and J. Sun, "Deep Residual Learning for Image Recognition," in 2016 IEEE Conference on Computer Vision and Pattern Recognition (CVPR), 2016, pp. 770-778.

[15] G. Huang, Z. Liu, L. V. D. Maaten, and K. Q. Weinberger, "Densely Connected Convolutional Networks," in 2017 IEEE Conference on Computer Vision and Pattern Recognition (CVPR), 2017, pp. 2261-2269.

[16] K. Xu, H. Huang, Y. Li, and G. Shi, "Multilayer Feature Fusion Network for Scene Classification in Remote Sensing," IEEE Geoscience and Remote Sensing Letters, vol. 17, no. 11, pp. 1894-1898, 2020.

[17] P. Jiang, B. Wan, Q. Wang, and J. Wu, "Fast and Efficient Facial Expression Recognition Using a Gabor Convolutional Network," IEEE Signal Processing Letters, vol. 27, pp. 1954-1958, 2020.

[18] X. Dong, H. Zhou, and J. Dong, "Texture Classification Using Pair-Wise Difference Pooling-Based Bilinear Convolutional Neural Networks," IEEE Transactions on Image Processing, vol. 29, pp. 8776-8790, 2020.

[19] L. d. F. Costa, F. A. Rodrigues, G. Travieso, and P. R. Villas Boas, "Characterization of complex networks: A survey of measurements," Advances in physics, vol. 56, no. 1, pp. 167-242, 2007.

[20] S. Boccaletti, V. Latora, Y. Moreno, M. Chavez, and D.-U. Hwang, "Complex networks: Structure and dynamics," Physics reports, vol. 424, no. 4-5, pp. 175-308, 2006.

[21] M. Varma and A. Zisserman, "A Statistical Approach to Material Classification Using Image Patch Exemplars," IEEE Transactions on Pattern Analysis and Machine Intelligence, vol. 31, no. 11, pp. 2032-2047, 2009.

[22] P. Mallikarjuna, A. Targhi, M. Fritz, E. Hayman, B. Caputo, and J. O. Eklundh, "THE KTH-TIPS2 database," 07/09 2006.

[23] T. Ojala, T. Maenpaa, M. Pietikainen, J. Viertola, J. Kyllonen, and S. Huovinen, "Outex - new framework for empirical evaluation of texture analysis algorithms," in Object recognition supported by user interaction for service robots, 2002, vol. 1, pp. 701-706 vol.1.

[24] Z. Baochang, G. Yongsheng, Z. Sanqiang, and L. Jianzhuang, "Local Derivative Pattern Versus Local Binary Pattern: Face Recognition With High-Order Local Pattern Descriptor," IEEE Transactions on Image Processing, vol. 19, no. 2, pp. 533-544, 2010.

[25] J. Zhang, J. Liang, and H. Zhao, "Local energy pattern for texture classification using self-adaptive quantization thresholds," IEEE transactions on image processing, vol. 22, no. 1, pp. 31-42, 2012.

[26] R. Mehta and K. Egiazarian, "Dominant Rotated Local Binary Patterns (DRLBP) for texture classification," Pattern Recognition Letters, vol. 71, pp. 16-22, 2016/02/01/ 2016.

[27] I. El khadiri, A. Chahi, Y. El merabet, Y. Ruichek, and R. Touahni, "Local directional ternary pattern: A New texture descriptor for texture classification," Computer Vision and Image Understanding, vol. 169, pp. 14-27, 2018/04/01/ 2018.

[28] Y. El merabet and Y. Ruichek, "Local Concave-and-Convex Micro-Structure Patterns for texture classification," Pattern Recognition, vol. 76, pp. 303-322, 2018/04/01/ 2018.

[29] M. Verma and B. Raman, "Local neighborhood difference pattern: A new feature descriptor for natural and texture image retrieval," Multimedia Tools and Applications, vol. 77, no. 10, pp. 11843-11866, 2018/05/01 2018.

[30] N. Alpaslan and K. Hanbay, "Multi-Scale Shape Index-Based Local Binary Patterns for Texture Classification," IEEE Signal Processing Letters, vol. 27, pp. 660-664, 2020.

[31] F. Chollet, "Xception: Deep Learning with Depthwise Separable Convolutions," in 2017 IEEE Conference on Computer Vision and Pattern Recognition (CVPR), 2017, pp. 1800-1807.

[32] B. Zoph, V. Vasudevan, J. Shlens, and Q. V. Le, "Learning Transferable Architectures for Scalable Image Recognition," in 2018 IEEE/CVF Conference on Computer Vision and Pattern Recognition, 2018, pp. 8697-8710.

[33] C. Sitaula, Y. Xiang, A. Basnet, S. Aryal, and X. Lu, "HDF: Hybrid Deep Features for Scene Image Representation," in 2020 International Joint Conference on Neural Networks (IJCNN), 2020, pp. 1-8.

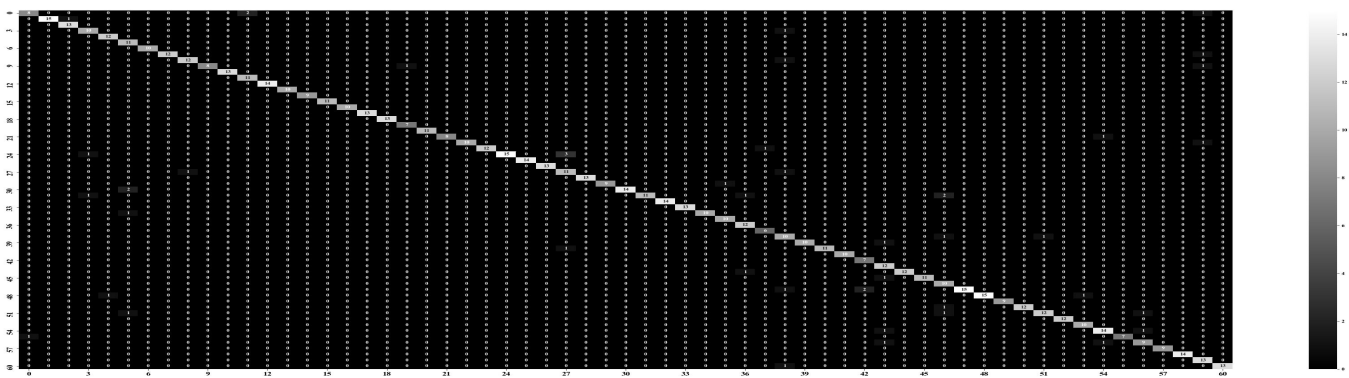

(a)

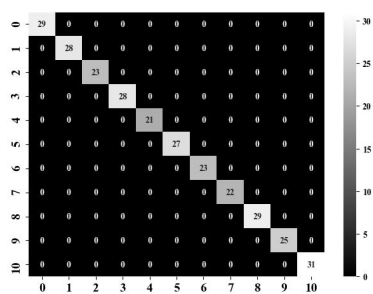

(b)

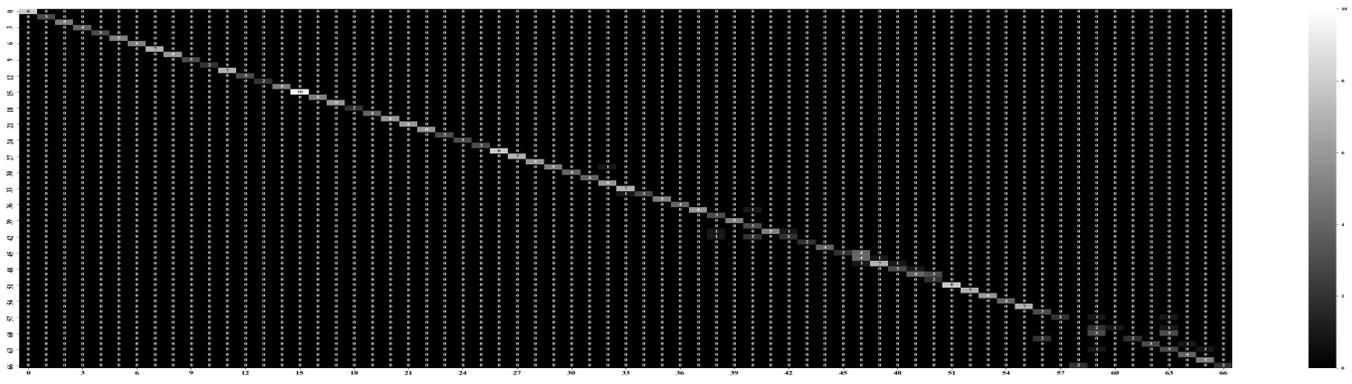

(c)

Fig. 4. Confusion matrices. (a) CUReT. (b) KTH-TIPS2-b. (c) Outex-TC-0001.